\documentclass[aps,pre,twocolumn,showpacs]{revtex4}
\usepackage{epsfig}
\usepackage{times}
\begin{document}

\title{Evolutionary prisoner's dilemma game on hierarchical lattices}
\author{Jeromos Vukov}
\affiliation {Department of Biological Physics, E\"otv\"os
University, H-1117 Budapest, P\'azm\'any P\'eter s\'et\'any 1/A,
Hungary}

\author{Gy\"orgy Szab\'o}
\affiliation {Research Institute for Technical Physics and
Materials Science, P.O.Box 49, H-1525 Budapest, Hungary}

\date{\today}
\begin{abstract}
An evolutionary prisoner's dilemma (PD) game is studied with
players located on a hierarchical structure of layered square
lattices. The players can follow two strategies [$D$ (defector)
and $C$ (cooperator)] and their income comes from PD games with
the ``neighbors.'' The adoption of one of the neighboring
strategies is allowed with a probability dependent on the payoff
difference. Monte Carlo simulations are performed to study how the
measure of cooperation is affected by the number of hierarchical
levels ($Q$) and by the temptation to defect. According to the
simulations the highest frequency of cooperation can be observed
at the top level if the number of hierarchical levels is low
($Q<4$). For larger $Q$, however, the highest frequency of
cooperators occurs in the middle layers. The four-level
hierarchical structure provides the highest average (total) income
for the whole community.
\end{abstract}


\maketitle

\section{\label{sec:intr}Introduction}

The evolutionary prisoner's dilemma game
(PDG)~\cite{maynard:82,hofbauer:98,weibull:95,hamilton:amnat63,axelrod:84}
is a very useful tool to examine the maintenance of cooperative
behavior among selfish individuals. Investigating this model, we
can study the behavior of a fictive society: we can collect data
about the change of the total income when varying the model
parameters, e.g., payoffs, set of strategies, topological
structure of interaction, rules controlling the choice of a new
strategy, noise, and external constraints.

The original PDG is a version of matrix games describing the
interaction between two players. The payoffs of the players depend
on their simultaneous decisions to cooperate or defect. In this
paper we use rescaled payoff parameters without any loss of
generality in the evolutionary PDGs~\cite{nowak:jbifchaos93}.
Thus, mutual cooperation yields the players unit income providing
them with the highest total payoff. On the contrary, their payoffs
are zero for mutual defection. If one of the players defects while
the other cooperates then the defector receives the highest
individual payoff ($b>1$), i.e., there is a temptation to defect,
while the cooperator's income (called sucker's payoff) is the
lowest one ($c<0$) . In the PDG the cooperator's loss goes beyond
the profit of the defector, i.e., $b+c<2$. The values of these
payoffs create an unresolvable dilemma for intelligent players who
wish to maximize their own income; namely, defection brings higher
individual income independently of the other player's decision but
for mutual defection they receive the second worst result.

In the case of evolutionary multiagent
systems~\cite{axelrod:84,blume:geb93} we face a totally changed
situation. The players' income is gained from iterated PDGs played
with different coplayers. In the present study we handle only the
two simplest strategies which are independent of the coplayers'
decisions: the first one is always cooperating, while the second
one always chooses defection. The players following these
strategies are called cooperators ($C$) and defectors ($D$)
respectively. In evolutionary games the players can adopt (learn)
one of the more successful strategies from the coplayers they
interact with. Generally, the probability of strategy adoption is
determined by the payoff difference.

The mean field approximation predicts that in a society of players
following the $C$ or $D$ strategy the cooperation dies out after a
short time. This happens in the case of the one-dimensional chain
if the interaction is limited to nearest neighbors. At the
boundary of a $C$ and $D$ cluster the defectors always have higher
payoff than the cooperators; thus cooperation vanishes quickly.

To demonstrate the advance of local interactions in the spatial
games, Nowak and May~\cite{nowak:jbifchaos93} have created a
two-dimensional cellular automaton where the players could follow
the $C$ or $D$ strategy. Their investigations revealed that the
cooperators overrun the territory of defectors along straight-line
fronts while the defectors' invasion can be seen along irregular
boundaries. It was shown that as a result of these invasion
processes the coexistence of $D$ and $C$ strategies took place
with a population ratio determined by the model parameters. Noisy
effects make the boundaries irregular and give more chance for
defection~\cite{nowak:jbifchaos94,szabo:pre98}. Randomly chosen
empty sites on the lattice further the maintenance of
cooperation~\cite{nowak:jbifchaos94,vainstein:pre01} by holding up
the spreading of defectors. These sites behave like defectors (for
$c=0$), but they cannot disperse; they are ``sterile.'' Finally,
it turned out that the short range interactions between the
localized players favor the maintenance of cooperation under some
conditions (e.g., for small $b$ values) even if just the simplest
strategies ($C$ or $D$) are allowed
\cite{nowak:jbifchaos93,nowak:jbifchaos94,acs:schweitzer,nakamaru:jtb97,szabo:pre98,
hauert:jtb02}. If $b$, the temptation to defect, exceeds a
threshold value depending on the evolutionary rules, then the
advantage of local interactions will not be enough to preserve
cooperation, so defection overcomes.

Recently spatial PDGs have been studied on different social
networks. First, different studies were performed to reveal what
happens on small-world networks~\cite{watts:nature98}. It is found
that cooperation can be maintained on these networks in a wide
range of parameters
\cite{abramson:pre01,kim:pre02,ebel:pre02,pre2004:szabovukov}.

Nowadays the research is concentrated on the revelation of those
circumstances for which the total income of the society can reach
its highest value. The other main purpose is to find networks
describing social systems more and more adequately.

First we studied the PDG on such a scale-free hierarchical
structure suggested by Ravasz and Barab\'asi
\cite{ravasz:barabasi} where the distribution of the clustering
coefficients is similar to that in social systems. The results of
this model are enclosed in the Appendix.

As the cooperation is not maintained in this system, we modified
the model's structure; we launched the study of a hierarchical
structure with many ``horizontal links.'' The equilibrium strategy
concentrations and average payoffs are analyzed as functions of
$b$ for each hierarchical level. By this means we can get some
information about those hierarchical structures providing the
highest total payoff for such a society.

\section{\label{sec:model}The model}

\begin{figure}[t]
\centerline{\epsfig{file=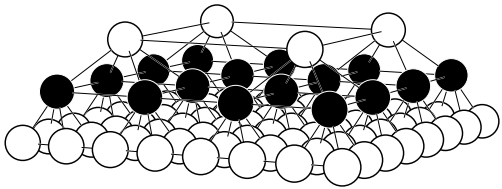,width=8cm}}
\caption{\label{fig:hier1d2d}Three-level ``hierarchical lattice''
constructed from square lattices (perspective view). Within a
level the players are linked to their nearest neighbors, and to
four other players located geometrically under them at the level
below. The horizontal periodic boundary conditions are not
indicated. White spheres indicate players at the first and the
third layers; meanwhile black spheres are for players at the
second level. In our notation the highest hierarchical level
($q=1$) is at the top.}
\end{figure}

We consider an evolutionary PDG with players located on the sites
of a hierarchical structure (lattice). The three-level version of
this structure is shown in Fig.~\ref{fig:hier1d2d}.

The $Q$-level hierarchical structure is constructed from the sites
of $Q$ square lattices positioned above each other; meanwhile, the
corresponding lattice constant is doubled level by level. These
levels are labeled by $q=1, \ldots , Q$ from top to bottom. Within
a given level the sites are linked (horizontally) to their four
nearest neighbors. The undesired edge effects are eliminated by
applying horizontal periodic boundary conditions for each level.
In addition, the sites of the level $q$ ($q=2, \ldots , Q$) are
divided into $2\times 2$ blocks whose four sites are linked to a
common site above belonging to the level $q-1$. This means that
for such a connectivity structure each player has four additional
links to its ``staff members'' and another one to its ``chief.''
Exceptions are naturally the players at the top ($q=1$) and bottom
($q=Q$) levels where the ``chiefs'' and ``staff members'' are
missing.

Consequently, the players at the top, middle, and bottom levels
have eight, nine, and five neighbors, and the corresponding values
of the clustering coefficient \cite{bollobas:98} are $1/7$, $1/6$,
and $1/5$, respectively.

Note that this structure preserves the (horizontal) translational
and rotational symmetry of the square lattice at the top level.
For technical convenience, the horizontal sizes of the bottom
layer are chosen to be $L=2^k$ ($k>Q$) with a unit lattice
constant. In this case this structure contains
$N=4[4^k-4^{(k-Q)}]/3$ sites (players).

The players follow one of the above mentioned $C$ or $D$
strategies. The spatial distribution of strategies is described by
a two-dimensional unit vector for each site $x$, namely,
\begin{equation}
s_x=\left(
\begin{array}{c}
1\\
0 \end{array} \right) \;\; \mbox{or} \;\; \left(
\begin{array}{c}
0\\
1 \end{array} \right)
\label{eq:states}
\end{equation}
for defectors and cooperators, respectively. Each player plays a
PDG with its ``neighbors'' (coplayers) defined by the above
structure and the incomes are summed. The total income of the
player at the site $x$ can be expressed as
\begin{equation}
M_x=\sum_{y \in \Omega_x}s_x^T  A  s_y \label{eq:mtotx}
\end{equation}
where the sum runs over all the neighboring sites of $x$ (this set
is indicated by $\Omega_x$) and the payoff matrix has a rescaled
form:
\begin{equation}
A=\left(
\begin{array}{cc}
$0$ & $b$ \\
$c$ & $1$ \\
\end{array}
\right), \label{eq:payoff}
\end{equation}
where $1<b<2-c$ and $c<0$ for the present PDGs. Since the work by
Nowak and May \cite{nowak:jbifchaos93} the parameter $c$ is
usually fixed to zero; therefore in our analysis we use the same
value.

In the evolutionary games the players are allowed to adopt the
strategy of one of their more successful neighbors. In the present
work the success is measured by the ratio of total individual
income and the number of neighbors (games), i.e.,
\begin{equation}
m_x=\frac{M_x}{|\Omega_x|} \;,
\label{eq:mx}
\end{equation}
where $|\Omega_x|$ indicates the number of neighboring players at
site $x$. This choice suppresses the advance coming from the
larger number of neighbors.

In the present evolutionary procedure the randomly chosen player
($x$) can adopt one of the (randomly chosen) coplayer's ($y$)
strategy with a probability depending on the difference of
normalized payoff ($m_x-m_y$) as
\begin{equation}
W(s_x\leftarrow s_y) = {1 \over 1 + \exp {[(m_x-m_y)/T]} } \;,
\label{eq:w}
\end{equation}
where $T$ indicates the noise \cite{szabo:pre98,blume:geb93}. This
definition of $W$ involves different effects (fluctuations in
payoffs, errors in decision, individual trials, etc.). Henceforth,
we consider the effects of $Q$ and $b$ on the measure of
cooperation for $T=0.02$.

\section{\label{sec:evol}Average strategy frequencies and payoffs}

The Monte Carlo (MC) simulations are carried out by varying the
values of $b$ and $Q$ for fixed $c$ and $T$ values as mentioned
above. For small system size, the MC simulations end up in one of
the homogeneous absorbing states (homogeneous $C$ or $D$ phase)
within a short time. In order to avoid this undesired phenomenon
we have investigated sufficiently large systems (containing $N
\simeq 10^5 - 10^6$ players) where the amplitudes of population
fluctuations are considerably smaller than the corresponding
average values.

\begin{figure}[t]
\centerline{\epsfig{file=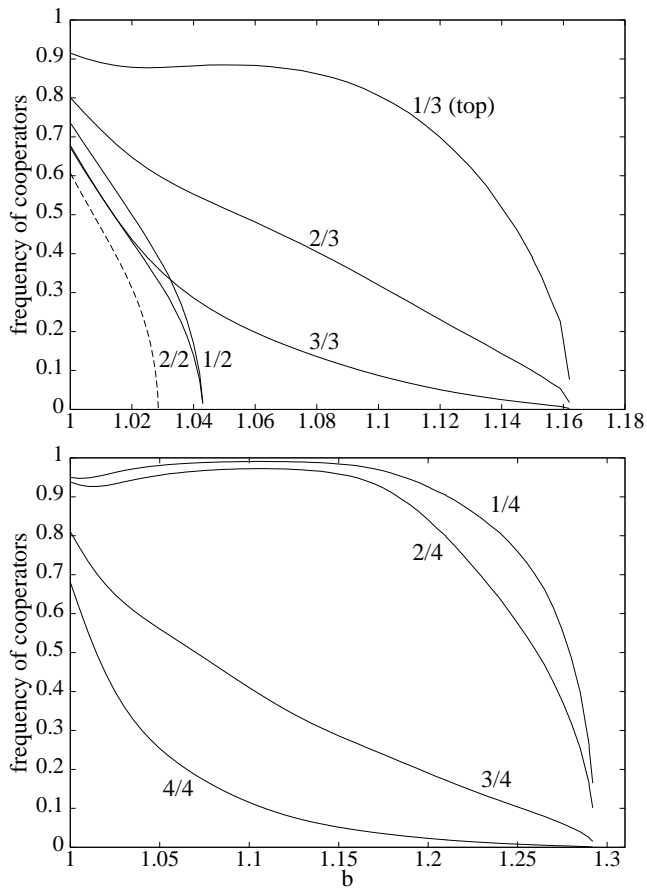,width=8.5cm}}
\caption{\label{fig:2dh234mcsb}Average frequency of cooperators as
a function of $b$ at different $q$ levels for $Q=2,3$ (top), and 4
(bottom). The dashed line (top) indicates the results obtained on
the square lattice ($q=Q=1$).}
\end{figure}
\begin{figure}[t]
\centerline{\epsfig{file=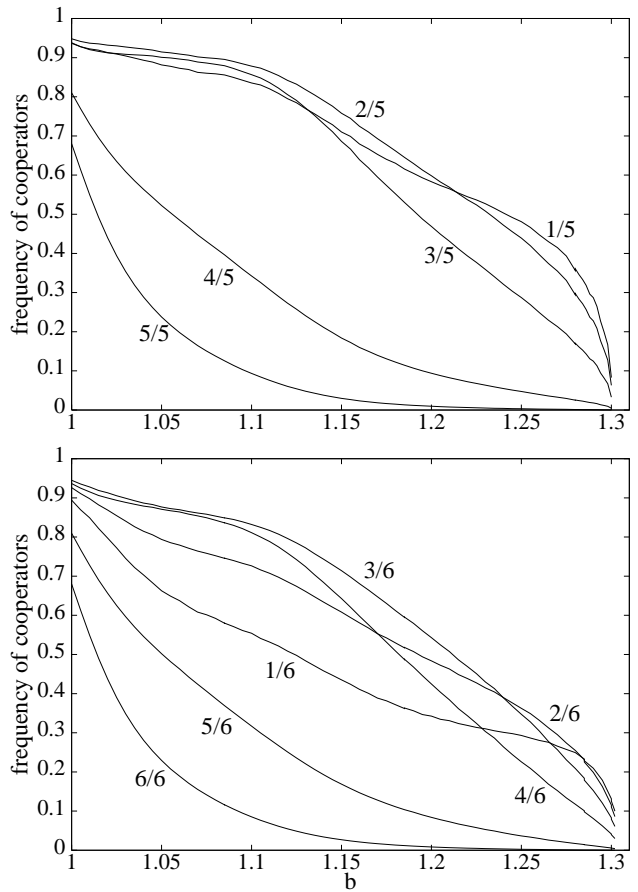,width=8.5cm}}
\caption{\label{fig:2dh56mcsb}The frequency of cooperators vs $b$
on all the hierarchical levels for $Q=5$ (top) and 6 (bottom).}
\end{figure}

The MC simulations are started from a random initial distribution
of strategies where the average frequency of both strategies is
1/2. The above described evolutionary steps (strategy adoptions)
are repeated until the system reaches the steady state we study.
It is checked in some cases that the steady states are independent
of the initial conditions. The steady states are reached after
some transient time varied from $2000$ to $50$ $000$ Monte Carlo
steps (MCS), where during the time unit 1 MCS each player has a
chance once on average to modify its strategy.

In the course of the simulations we have recorded the current
frequencies [$\rho_D(t)$ and $\rho_C(t)=1-\rho_D(t)$] and payoffs
[$m_D(t)$ and $m_C(t)$] as a function of time $t$ for both
strategies at each level. The corresponding average values are
obtained by averaging through $16$ $000$ MCS. With knowledge of
these quantities, we can determine other quantities concerning the
whole system (total income of the society, strategy distributions
referring to the whole system, etc.).

First we study how the strategy populations at each level depend
on $Q$ and $b$. The results for $Q=2,3$, and $4$ are plotted in
Fig.~\ref{fig:2dh234mcsb}, and Fig.~\ref{fig:2dh56mcsb} represents
the data for $Q=5$ and $6$. The corresponding system sizes are
$N=81$ $920$, $344$ $064$, $348$ $160$, $349$ $184$, and $349$
$440$. For these sizes the statistical error is comparable with
the line thickness. It is observed that there always exists a
threshold value ($b_{c}$) for any value of $Q$. For $b>b_c$ the
cooperators become extinct. This critical value, however, depends
on the number of hierarchical levels: $b_{c}(Q)$. The exact value
of $b_c$ for each $Q$ is not determined because the thermalization
time (as well as the fluctuation) grows very fast when approaching
the critical points. Consequently, the precise determination of
$b_c$ requires a very long computer time (this is the reason why
the end points are not represented in the figures). It is expected
that the extinction of cooperators at large spatial scales belong
to the two-dimensional directed percolation universality class
\cite{szabo:pre98,hinrichsen:ap00}. Unfortunately, the numerical
justification of this conjecture exceeds our computer capacity.

We have performed MC simulations on the square lattice with
periodic boundary conditions ($Q=1$), too. The frequency of
cooperators vanishes continuously as a function of $b$ (dashed
line in the upper plot of Fig.~\ref{fig:2dh234mcsb}). The value of
the critical point is $b_c=1.028$ $524(5)$. The phase transition
belongs to the two-dimensional directed percolation universality
class~\cite{szabo:pre98,hinrichsen:ap00}.

Comparing the results of the two-level case with those on the
square lattice one can see an increase of $b_c$ that may have come
from the enhancement in the average number of neighbors (from $4$
to $5.6$) and/or the average clustering coefficient (from $0$ to
$0.189$). It is worth mentioning that both of these quantities
increase monotonically with $Q$. In the limit $Q \to \infty$ the
average number of neighbors goes to six and the average clustering
coefficient tends to $23/120=0.191\dot{6}$.

Figures \ref{fig:2dh234mcsb} and \ref{fig:2dh56mcsb} show clearly
that the frequencies of cooperators differ level by level. At the
same time, the phase transitions to the absorbing (homogeneous)
$D$ state occur for each level at the same critical value of $b$
dependent on $Q$. This feature is related to the fact that the
successful colonies of cooperators can pass their strategy to any
level although its probability depends on $q$.

Another striking message of the above numerical results is that
the lowest measure of cooperation always occurs at the bottom
level. The label of the most cooperative level, however, depends
on $Q$ and $b$. For $Q=2, 3$, and $4$ the frequency of cooperators
is the highest at the top level, and it is decreasing
monotonically when going downward on these hierarchical
structures. The simulations for the given parameters show
different behaviors if $Q > 4$. In the five-level system the
highest measure of cooperation is reached by players in the second
level in a wide range of $b$ and their frequency is exceeded by
the cooperators of the top level only in the vicinity of the
critical point. The interval where the measure of cooperation is
the highest at the top level becomes narrower for $Q=6$. In this
case one can observe two additional regions of $b$ where the
second and third levels exhibit the highest cooperativity.
Furthermore, within a wide range of $b$ the measure of cooperation
at the fourth level is higher than it is at the top level.

\begin{figure}[t]
\centerline{\epsfig{file=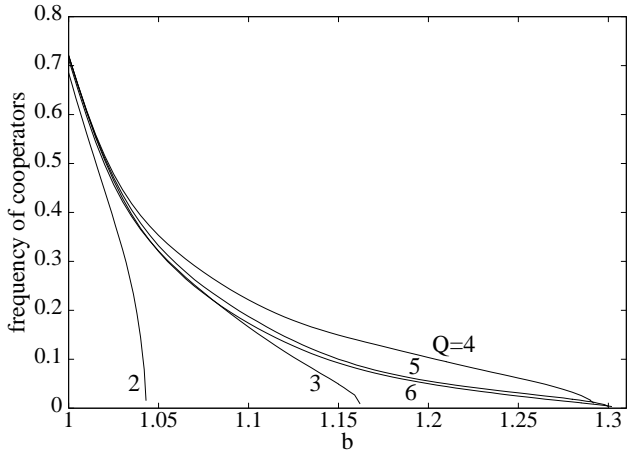,width=8.5cm}}
\caption{\label{fig:avc2dh}The frequency of cooperators in the
whole system as a function of $b$ and $Q$.}
\end{figure}
\begin{figure}[t]
\centerline{\epsfig{file=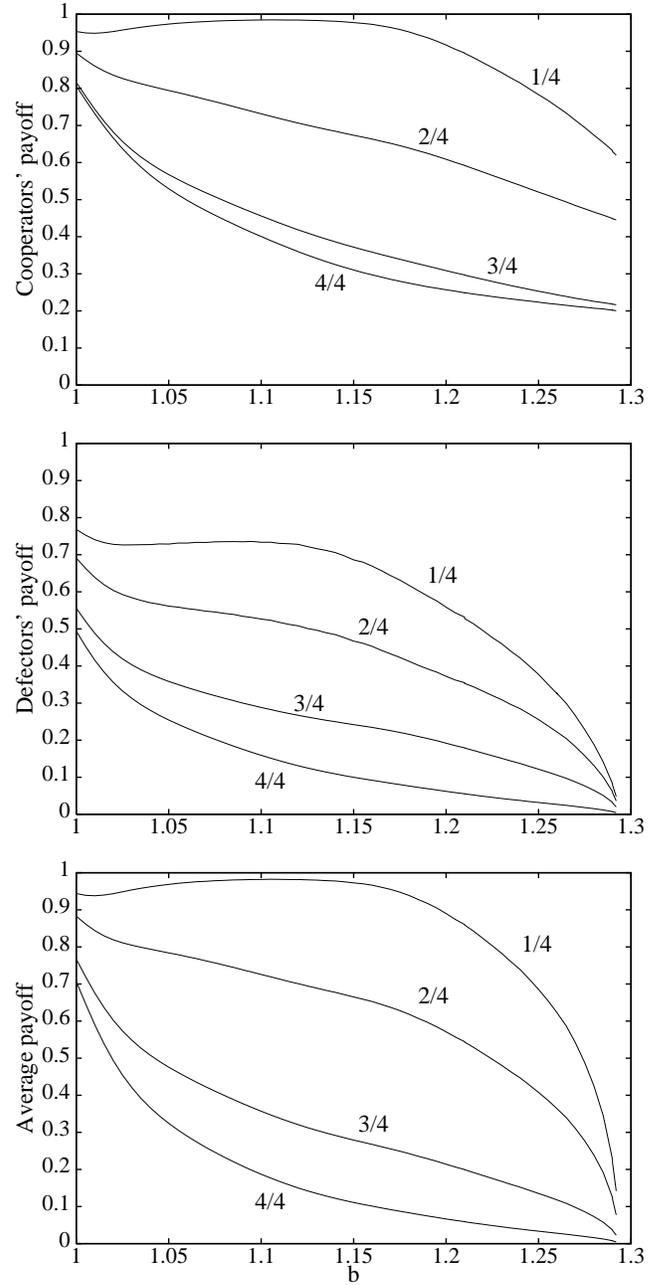,width=8.5cm}}
\caption{\label{fig:osszpo34}The average (normalized) payoffs as a
function of $b$ at each hierarchical level for $Q=4$. The top
(middle) plot shows the average payoffs of cooperators (defectors)
while the bottom figure indicates the weighted average payoffs at
each level.}
\end{figure}

Summarizing the data of the different levels, we can determine the
measure of cooperation in the whole system as a function of $Q$
and $b$. The average values are very low. The reason for this is
that the measure of cooperation is the lowest at the bottom level
where most of the players are located. As plotted in
Fig.~\ref{fig:avc2dh} the measure of cooperation has its maximum
for $Q=4$ for almost any value of $b$. It is observable that the
$b_c(Q)$ values increase monotonically with $Q$, and they seem to
converge to a value near $b_c(Q=6)$. This tendency is likely to
come from the mentioned convergence of the average number of
neighbors and/or the average clustering coefficient.

\begin{figure}
\centerline{\epsfig{file=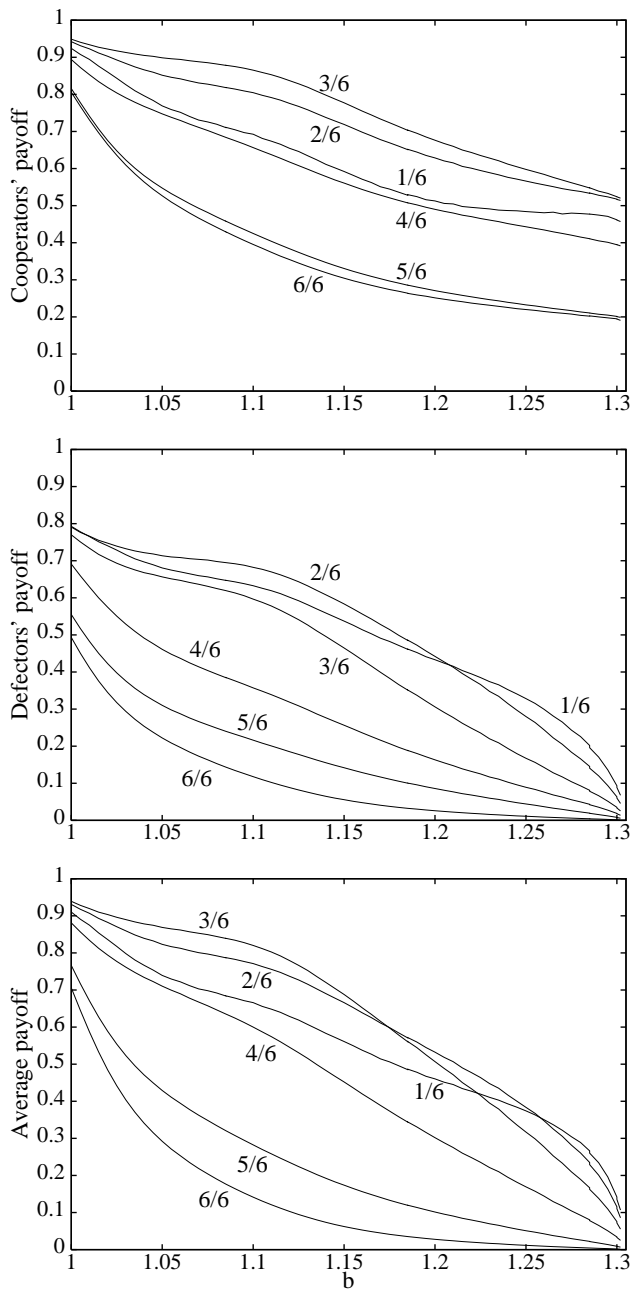,width=8.5cm}}
\caption{\label{fig:osszpo56}The average (normalized) payoffs vs
$b$ at each hierarchical level for $Q=6$.}
\end{figure}

Now we are going to analyze the average payoffs separately at the
different levels as well as in the whole system.
Figures~\ref{fig:osszpo34} and \ref{fig:osszpo56} show the average
payoffs of cooperators and defectors at each level for several $Q$
values. Beyond that, the players' average payoff at each level is
also presented. Comparing these figures with those representing
the frequency of cooperators for the same $Q$, one can observe
high similarity. This similarity is due to the fact that the
cooperative partners ensure (positive) income for both the
cooperators and defectors. (The total payoff for a player
following the $C$ or $D$ strategy and having $n$ neighbors
following the $C$ strategy is equal to $n$ or $b n$,
respectively.) Notice, furthermore, that the dominant part of the
total income is received by the cooperators whose average incomes
exceed those of defectors at each level. This fact comes from the
formation of colonies by cooperators explaining why their average
income remains positive for their vanishing concentrations.
Although along the boundary of these colonies the defectors
exploit cooperators, their average income is lower because most of
them are surrounded by defectors and receive zero income.
Evidently, the average payoff in the whole system (as well as the
defectors' payoffs for each level) disappears continuously when
approaching the critical point ($b \to b_c$) from below.

In the six-level system, the rank of the average defector payoff
differs from that of the cooperator payoff. The average income of
defectors is the highest at the second and the first levels. The
reason for this is that for $Q=6$, the highest measure of
cooperation is found at the middle levels and as the result of the
geometrical arrangement of the structure, these cooperators can be
exploited the most by the defectors located geometrically above
them.

For $Q=2$ and 3 the average payoffs exhibit similar features as
are found for $Q=4$ (see Fig.~\ref{fig:osszpo34}) while for $Q=5$
the system behavior is analogous to the case of the six-level
system.

For the different $Q$ values the $b$ dependence of the total
payoff per game (in the whole system) is very similar to the
measure of cooperation plotted in Fig.~\ref{fig:avc2dh}. This
means that the four-level hierarchy provides the highest average
income to the players for such connectivity structures.

\section{\label{sec:sum}Summary}

This work is devoted to studying the effect of the hierarchical
structure on the measure of cooperation in a society where the
pairwise interaction between the members is modeled by an
evolutionary PDG with two strategies ($C$ and $D$).

Using MC simulations we have determined the frequency of
cooperators as a function of $b$ (the measure of temptation to
choose defection) at the different hierarchical levels in the
$Q$-level structures. These systems exhibit a continuous
(critical) transition from the $C+D$ state to the homogeneous $D$
state if $b$ tends to $b_c$ depending on $Q$. During this
transition the cooperators become extinct simultaneously at each
level of the hierarchical structures.

The lowest measure of cooperation is always found at the bottom
level. In the close vicinity of the transition point the frequency
of cooperators is increasing gradually when going upward on the
hierarchical structures. For lower $b$ values, however,
significantly different behavior is observed if $Q>4$; namely, the
highest frequency of cooperators is found in the middle levels;
meanwhile the total payoff per game decreases if $Q$ is increased
for a fixed $b$. Surprisingly, the present simulations indicate
clearly the existence of an optimum number of hierarchical levels
where such a community can reach the highest income. For the
present hierarchical lattices the four-level structure provides
the highest income (productivity).

Another important message of the present investigations is related
to the importance of ``horizontal lattice structure'' (or the
additional horizontal links) in the hierarchical structures. As is
mentioned in the Introduction, the cooperators die out
exponentially in a similar evolutionary PDG model if the players
are located on a complex hierarchical network suggested by Ravasz
and Barab\'asi~\cite{ravasz:barabasi}. We think that further
systematic research is required to clarify the relationship
between the maintenance of cooperation and the topological
structure of connectivity.

\begin{acknowledgments}
This work was supported by the Hungarian National Research Fund
under Grant No. T-47003.
\end{acknowledgments}

\appendix*
\section{\label{sec:app}Evolutionary prisoner's dilemma game on
a scale-free hierarchical network}

In this model, the players are located on a scale-free network,
where the distribution of the clustering coefficients is similar
to that in social systems. The dynamical rule in this model is the
same as described in Sec.~\ref{sec:model}; the difference occurs
in the structure of the network.

\begin{figure}
\centerline{\epsfig{file=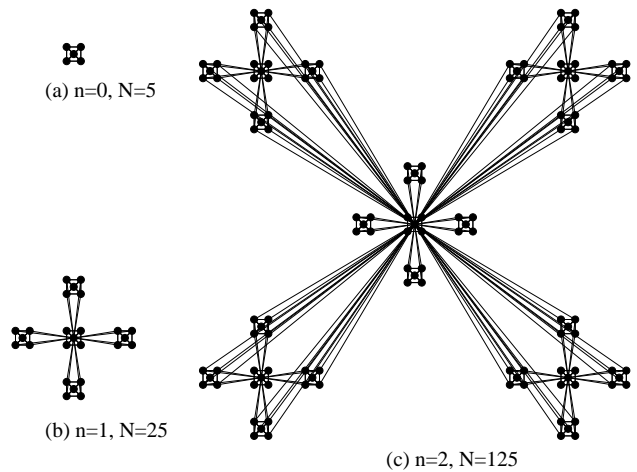,width=8.5cm}}
\caption{\label{fig:hierkep}Iterative steps (indicated by $n$) in
the construction of the scale-free network.}
\end{figure}

The construction of the network suggested by Ravasz and
Barab\'asi~\cite{ravasz:barabasi} is the following. In the first
step, there is a small cluster of five densely linked nodes
[Fig.~\ref{fig:hierkep}(a)]. In the next step, we generate four
replicas of this cluster and connect the four external nodes of
the replicated clusters to the central node of the old cluster,
creating a large 25-node module [Fig.~\ref{fig:hierkep}(b)].
Continuing the process, we again generate four replicas of this
25-node cluster, and connect the 16 peripheral nodes of each
replica to the central node of the old module
[Fig.~\ref{fig:hierkep}(c)], obtaining a new module of 125 nodes.
These replication and connection steps can be iterated
indefinitely; in each step the number of nodes is multiplied by a
factor of 5.

The generated graph has a power-law degree distribution with a
degree exponent $\gamma = 1+\ln 5 / \ln 4= 2.161$. In this network
for $N\ge 125$ the average clustering coefficient is $C\simeq
0.743$. The network's hierarchical structure is apparent: there
are several, small, fully connected five-node graphs, which are
clustered into larger 25-node graphs. The 25-node graphs are
clearly separated from each other, and they are clustered into
much larger 125-node graphs, etc.

The hierarchical levels are introduced in the following way. Being
in the center of bigger cluster means being at higher level in the
hierarchy. Figure~\ref{fig:hierkep}(b) shows a three-level
hierarchy: there is only one player at the first (highest)
hierarchical level, the one in the center of the entire structure;
the center nodes of the four ``replicas'' are at the second level,
while the other nodes are at the third (lowest) level (there are
20 nodes of this type). One iterative step increases the number of
hierarchical levels by 1. The analyzed network has $H=7$
hierarchical levels; therefore the number of players (nodes) is
$N=5^{H-1}=5^6=15$ $625$.

We have performed Monte Carlo simulations as described previously
by varying $b$ for fixed temperature ($T=0.01, 0.02$, and $0.05$),
to analyze what happens in a society where the players can follow
$C$ or $D$ strategies.

Two different stages can be distinguished as the system evolves.
The first one is a transient process, where most of the small
groups of cooperators die out. After this period ($\simeq 4000$
MCS), the cooperators remain on such groups of sites where they
can defend themselves against the defectors' ``attacks.'' In this
state the frequency of cooperators rather depends on the initial
distribution of strategies than the value of $b$.

The small groups of cooperators can survive in two types of
configurations. In the first case, they occupy a five-node
subgraph [see the graph in Fig.~\ref{fig:hierkep}(a)] with central
node at the level $H-1$. The other basic unit of the $C$ groups is
composed of the four peripheral nodes of a five-node graph with a
central node located at a hierarchical level $h\ge H-1$.

Within such cliques~\cite{bollobas:98}, if all the players
cooperate, then (due to the many internal links) they receive such
a high income from each other that it provides protection against
the external $D$ invaders even for the highest $b$ value. For this
particular structure the normalized payoff of the attacking
defectors is reduced drastically by the many neighboring $D$
strategies. Occasionally, however, these groups can be invaded by
defectors in the presence of noise. If a single $D$ strategy gets
into a clique then its offspring invade the whole group within a
short time because these internal defectors (who have only a few
defector neighbors) exploit all the internal cooperators.

\begin{figure}
\centerline{\epsfig{file=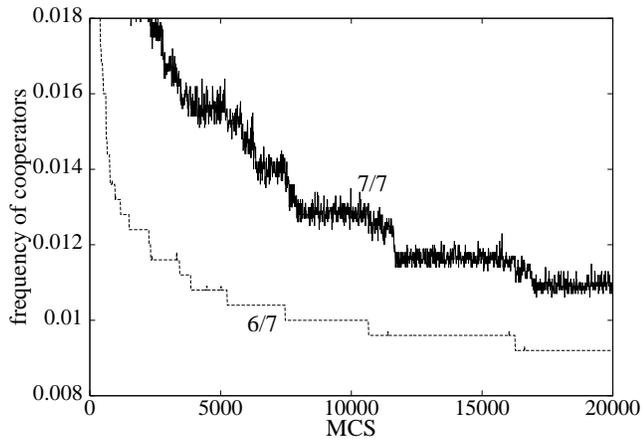,width=8.5cm}}
\caption{\label{fig:hlepcso}The frequency of cooperators as a
function of time at the two lowest hierarchical levels.}
\end{figure}

The above process implies a decrease in the number of cooperators'
cliques as well as in the average frequency of cooperators.
Figure~\ref{fig:hlepcso} shows a noisy steplike decrease of $C$
frequency in time where the magnitude of the steplike decrease
corresponds to the disappearance of a cooperator clique described
above. As the players in the small cooperator groups have higher
average payoffs than the defectors, they invade continuously the
territory of the defectors. At the same time, due to the network's
topology, the mutual cooperation cannot persist within these
territories; therefore the $D$ strategy will recapture them in a
short time.

As a result of the above process the system will end up in the
absorbing $D$ state for any $b$. The average lifetime of the rare
cooperator cliques depends on $b$ and $T$. More precisely, the
extinction of the cooperators becomes slower for lower $b$ and
$T$. The probability of the first successful $D$ attack against a
$C$ clique can be estimated by the application of Eq.~\ref{eq:w},
and the corresponding theoretical predictions are consistent with
the results of the simulations.

\rule{0cm}{0cm}

\rule{0cm}{0cm}

\rule{0cm}{0cm}

\rule{0cm}{0cm}

\rule{0cm}{0cm}


\end{document}